\documentstyle[12pt]{article}
\begin{document} 
\vspace*{-1in} 
\renewcommand{\thefootnote}{\fnsymbol{footnote}} 
\vskip 65pt 
\begin{center} 
{\Large \bf A new proposal for glueball exploration in Hard Gluon
Fragmentation}\\
\vspace{8mm} 
{\bf 
Probir Roy\footnote{probir@theory.tifr.res.in}, 
K.~Sridhar\footnote{sridhar@theory.tifr.res.in}
}\\ 
\vspace{10pt} 
{\sf  Department of Theoretical Physics, Tata Institute of 
Fundamental Research,\\  
Homi Bhabha Road, Bombay 400 005, India. } 
 
\vspace{80pt} 
{\bf ABSTRACT} 
\end{center} 
\vskip12pt 
An unambiguous identification of glueballs in experiments will be of great 
significance, because their existence is an important test of QCD.
The proposal, advanced here, is to experimentally search for glueballs
as peaks
in the invariant mass of a leading $K_S$-pair fragmenting from an energetic 
gluon jet
out of high-statistics three-jet events in hadronic decays of the weak
neutral $Z$ boson. Using a physically motivated model of the gluon-glueball
fragmentation function, we find a substantial fragmentation
rate into a leading glueball. It is very likely that a search,
along the lines suggested here by any of the four groups at the
Large Electron Positron collider at CERN, will prove fruitful.
\setcounter{footnote}{0} 
\renewcommand{\thefootnote}{\arabic{footnote}} 
 
\vfill 
\clearpage 
\setcounter{page}{1} 
\pagestyle{plain}
\noindent Gluons, the confined colour-octet mediators of strong interactions
in quantum chromodynamics (QCD), `shine \cite{close1} in their own light'.
They have self-interactions as a consequence of non-abelian gauge symmetry.
At short distances $<< 1 {\rm GeV}^{-1}$, their coupling strength decreases
via renormalisation group evolution to yield an asymptotically free 
\cite{peskin} weak-coupling description. But at distances $\ge (\Lambda_{\rm
QCD})^{-1}$, where $\Lambda_{QCD}$ is the QCD scale $\sim 200$~MeV, the 
coupling strength increases to a strong enough value to cause colour 
confinement. With such strong couplings, colour-singlet gluonic bound states
or {\it glueballs} \cite{close1} are expected to form. Indeed, there exist
strong theoretical arguments \cite{west} favouring such formation.

Simple representations of scalar, pseudoscalar and tensor glueball
fields are $G(x) \sim {\rm Tr} F_{\mu\nu}(x) F^{\mu\nu}(x)$,
$\tilde G(x) \sim {\rm Tr} F_{\mu\nu}(x) \tilde F^{\mu\nu}(x)$,
and $G_{\mu}^{\nu}(x) \sim {\rm Tr} F_{\mu\rho}(x) F^{\rho\nu}(x)$,
respectively. Here $F_{\mu\nu}(x)$ is the covariant colour-contracted
gluon field-strength tensor in standard notation \cite{peskin} and
$\tilde F^{\mu\nu}(x)$ is its dual. Theoretical studies, carried out
over two decades, suggest the existence of glueball states in the
few GeV mass-range. Such considerations cover bag \cite{bag},
quasiparticle \cite{quasi} and instanton \cite{instanton} models and even those
on supergravity \cite{sugra}. There have also been glueball simulations
\cite{lattice} on the lattice which fall in the same ambit.

All of the above studies predict the lightest member of the glueball
spectrum to be a scalar. In fact, the lattice approach \cite{lattice} 
pins down its mass within the window 1.5-1.7 GeV. A scalar mass of
1.5~GeV or so is suggested anyway from the square-root of the inverse
of the slope $\simeq 0.4/{\rm GeV}^2$ of the Pomeron trajectory, describing
high-energy diffraction, if the latter is identified as the grandparent
of the scalar glueball trajectory. However, there is controversy over
the predicted spin, parity of the lightest glueball. Estimates \cite{sumrules}
from QCD sum-rules show preference for the latter being a pseudo-scalar, while
some field-theoretic models \cite{schaden} suggest that it could be tensor. 
Glueballs will, of course, be unstable against hadronic decay. But, on
account of the $\sqrt{OZI}$ rule \cite{close1}, their width should not
be much more than 100 MeV or so. Thus they are expected to be narrower
than typical $q \bar q$ resonances in that mass range, though this
characteristic feature may get diluted due to glueball-meson mixing.

Since glueballs are inherently quantum chromodynamic in nature, the
confirmation of a glueball would constitute direct evidence for QCD. Much
effort, as reviewed in Refs.~\cite{close1, pennington, close2}, has gone into
the production and detection of such states. First of all, glueballs
need glue-rich production channels. They are scarcely produced in
usual quark-antiquark creation, annihilation and rearrangement subprocesses.
A further complication is that \cite{farrar} glueballs are expected to mix 
significantly
with flavour-singlet $q \bar q$ mesons of the same spin and parity. For 
instance, the central region of hadroproduction is characterised by 
the $gg$ production channel. Nevertheless, careful filtering procedures
\cite{kirk}
have to be devised to avoid misidentifying flavour singlet mesons as
glueballs among resonances produced here. There have been several quests
in this direction \cite{search1}. Another probe \cite{search2, search3}
has been the radiative decay of charmonium ($J/\psi$) where
the photon could recoil against a glueball. However, the reduced statistics
of a radiative process constitute a limiting factor here. Several interesting
candidates have emerged from both of these studies~: $f_0 (1500)$, $f_J 
(1710)$ with $J=0,2$, $\xi (2230)$ etc. The glueball interpretation of
these flavour singlet mesons is quite plausible. Still, statistically
significant clinching evidence for a conclusive glueball identification 
has been lacking so far and alternative avenues need to be explored. This
motivates us to propose a new way of gathering such evidence.

\begin{figure}[ht]
\vskip 3.5in\relax\noindent
\includegraphics{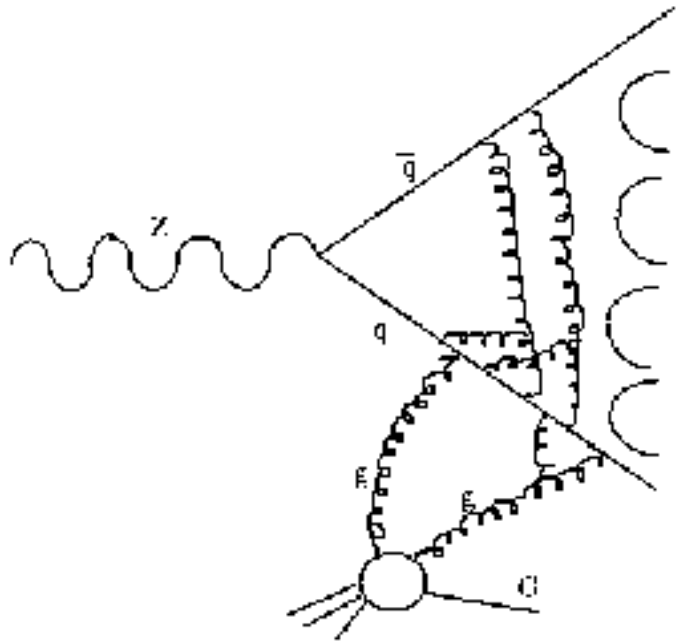} 
\vskip 0.25in\relax\noindent
\caption{Glueball fragmentation from a hard gluon in $Z \rightarrow q \bar q g$
decay}
\end{figure}

We are guided by the simple ides that a sufficiently hard gluon, hadronizing 
as an energetic jet, will naturally fragment first into a leading glueball.
Because of the reasonably large mass of the glueball, a sizable amount of
rapidity will be taken up in the process of lifting a glueball from the
vacuum, leaving the residual gluon as quite soft (Fig.~1). Such a possibility
was first mooted \cite{probir} two decades ago in the context of collinear
hadronic decays of a heavy quarkonium. Subsequent experimental searches
on the $\Upsilon$ resonance were unsuccessful, but then the three gluons
emerging from the bound $b \bar b$ annihilation in hadronic $\Upsilon$
decay are not sufficiently energetic to form isolated jets, which are
necessary for the fragmentation process. With higher energy gluon jets,
fragmentation processes become important as is evidenced by the study
of $J/\psi$ production $via$ gluon fragmentation at the Tevatron \cite{jpsi}. 
In fact, at high energy colliders gluon fragmentation becomes dominant and
becomes an important discovery channel for new particles. We are thus 
naturally led to direct our attention to hard, isolated gluon jets in
the sample of $Z\rightarrow q \bar q g$ three-jet events at LEP. The
least energetic of the three jets in the sample is taken, with a high
degree of reliability (with an efficiency of about 70\%), 
to be a gluon jet. Indeed, this is what is borne
out in simulations \cite{simulate} based on perturbative QCD. Even after the
imposition of a cut of $E_{\rm jet} > 15$~GeV in the $Z$ rest-frame, one
should still be left with nearly hundred thousand events from the LEP1 $Z$
sample. This is a rich repository of events containing isolated hard 
gluon jets. One is likely to get an observational handle on any glueball
produced in them if one can estimate the gluon-glueball fragmentation
probability, multiplied by the branching ratio for the glueball decaying
into a $K_S$-pair, that is credible even within an order of magnitude.
The glueball will show up as a peak in the $K_S K_S$ invariant mass 
spectrum, studied in the gluon (but not in the $q-$ or $\bar q -$) jet. For a 
scalar or pseudoscalar glueball, no correlations are expected between
the $K_S$-directions and the jet axis; for a tensor one there will, in 
general, be such correlations.

Let us quantitatively consider the question of the glueball fragmentation
of a hard gluon in the three-jet final state of $Z$ hadronic decay at
LEP. We will consider the quantity $\Gamma (Z \rightarrow q \bar q G X)$,
the partial width for the $Z$ to decay into a $q \bar q$ pair plus a
glueball state, $G$, and other soft gluons. These soft gluons will
produce soft hadrons which in this inclusive mode we denote by $X$.
The hardness of the fragmenting gluon can be ensured by a cut on the 
energy of the gluon jet in the rest frame of the $Z$. It is expedient to write $\Gamma (Z \rightarrow q \bar 
q G X)$, in terms of the $q \bar q$ partial width, $\Gamma (Z \rightarrow q \bar
q)$. This can be done using the well-known expressions for $\Gamma (Z 
\rightarrow q \bar q g)$ in terms of $\Gamma (Z \rightarrow q \bar q)$,
given as
\begin{equation}
{d\Gamma(Z\rightarrow q \bar q g) \over dx_2 dx_3}= {2 \alpha_s \over 3 \pi}
\Gamma(Z\rightarrow q \bar q ) 
{x_1^2+x_2^2 \over (1-x_1) (1-x_2)} ,
\end{equation}
where $x_i= 2E_i/M_Z$ with $E_i=1,2,3$ denoting
the energy (in the $Z$ rest frame) of the antiquark, quark and the gluon jet,
respectively. We note that $x_1+x_2+x_3=2$. Using this expression, we can
write down the corresponding expression for $\Gamma (Z \rightarrow q \bar q 
G X)$.
It is more convenient to write this width in terms of $z$, which is the
fraction of the parent gluon energy carried by the glueball, rather than in terms
of $x_3$. After this transformation of variables, we fold $\Gamma
(Z\rightarrow q \bar q g)$ with the fragmentation function $D(z,Q^2)$,
where $Q^2$ is the scale at which the fragmentation function is evaluated.
The resultant expression is 
\begin{equation}
\Gamma(Z\rightarrow q \bar q G X) = {2 \alpha_s \over 3 \pi}
\Gamma(Z\rightarrow q \bar q ) \int dx_2 \int dz
{x_3 \over z} {x_1^2+x_2^2 \over (1-x_1) (1-x_2)} D(z,Q^2),
\end{equation}
where the limits of integration in the above expression are chosen in a way
consistent with the experimental cuts to be specified in detail below.

To estimate the glueball production rate, we need to make an ansatz for
the glueball fragmentation function.  The simplest assumption is to
consider the fragmentation of a high energy gluon into a glueball as
being analogous to the fragmentation of a valence quark into a meson.
This may appear unusual at first sight.  We know that all quark
jets predominantly fragment into mesons whereas in most of the gluon
jets -- studied in three-jet samples in $e^+e^-$ machines at CM
energies far below the $Z$-mass -- the parent gluon first goes into a
$q\bar q$ pair which hadronize in terms of $\pi$'s, $\eta$'s, $\rho$'s
etc.  Unlike the former, which is a zeroth order process, the latter
is $O(\alpha_s)$ in the rate; but the large mass of the glueball makes
it impossible for such a gluon to effect a zeroth order fragmentation
into it.  Our claim is that, once a gluon is very energetic, as is the
case for the one emitted by the $Z$ via $Z \rightarrow q\bar q g$, it
will easily overcome this threshold effect; its fragmentation into a
glueball state would then become a `valence-like' process.  For such a
gluon, fragmentation via the transition first into a $q\bar q$ pair
would be comparatively down by an $O(\alpha_s)$ factor just as quark
fragmentation via gluon radiation is smaller as compared with the
direct fragmentation into a meson of a valence quark.

There is more justification for the above assumption. 
A calculation \cite{carlson}, based on QCD sumrules, of the exclusive 
distribution of gluons inside a glueball (i.e. the wavefunction) 
shows that the results are very similar to that of the meson
wavefunction. In fact, these calculations suggest a somewhat
larger normalisation for the glueball wavefunction and it
is possibly true of the inclusive fragmentation function too.
But without dwelling too much on these finer points,
let us point out that this ansatz for the
fragmentation function is being made with the idea of estimating 
the rate of glueball production via fragmentation at LEP2 energies.
The assumption we make allows us to make a rough estimate for the
number of glueball events we expect to see at LEP2. Taking this
number as given, we can then try to understand whether it is
feasible to attempt a search for the glueball state through
its decay into mesons. In that sense, we should take the numerical
results presented here as a rough guide to decide on what kind
of search strategies will be appropriate in the experimental
situation. Moreover, we also present results with a different fragmentation function
and study the effect on our results of varying this input. For the
pion fragmentation function, we use the parametrisation of Ref.~\cite{pion} 
(which is a $1-z$ distribution with the normalisation obtained from 
a fit to pion production data) and use this as the glueball 
fragmentation function at the input scale $\mu_0 \sim 2$~GeV. To take
into account the fact that the glueball mass is quite substantial,
we multiply this fragmentation function with a multiplicative
threshold factor $(1 - 4 M_G^2 /E_g ^2)$, where $M_G$ is the mass
of the glueball and $E_g$ is the lab-frame gluon energy. The fragmentation function 
is then evolved to the scale typical of the 
fragmenting gluon using Altarelli-Parisi evolution. In the evolution, 
we have neglected the
non-diagonal anomalous dimensions, since their effects
are sub-leading. When we vary the fragmentation function, we choose 
a $(1-z)^2$ distribution instead, but we normalise the distribution
in such a way that the integrated probability is the same as in the
case of the $1-z$ distribution.

In order to make contact with the experimental jet selection criteria
used in the LEP experiments, we require that the lowest energy parton
is identified as the gluon and that is the fragmenting parton. Also,
it is usual to select the jet sample by requiring a minimum cut, 
$d_{\rm min}$ on the quantities $d_{ij}$, defined as
\begin{equation}
d_{ij} = {2 E_i E_j {\rm sin} \theta_{ij} /2 \over E_i+E_j} ,
\end{equation}
where the $i, j$ indices refer to the three partons in the three-jet 
final state.
Following the experimental cuts, we take $d_{\rm min}$ to be 7~GeV.
In addition, we also require that the gluon energy be above a
minimum value, $E_{\rm cut}$. Since $\Gamma(Z\rightarrow q \bar q G X)$ 
is a function of $E_{\rm cut}$, we study this functional dependence 
by varying $E_{\rm cut}$. 

\begin{figure}[ht]
\vskip 0.5in\relax\noindent
          \relax{\includegraphics{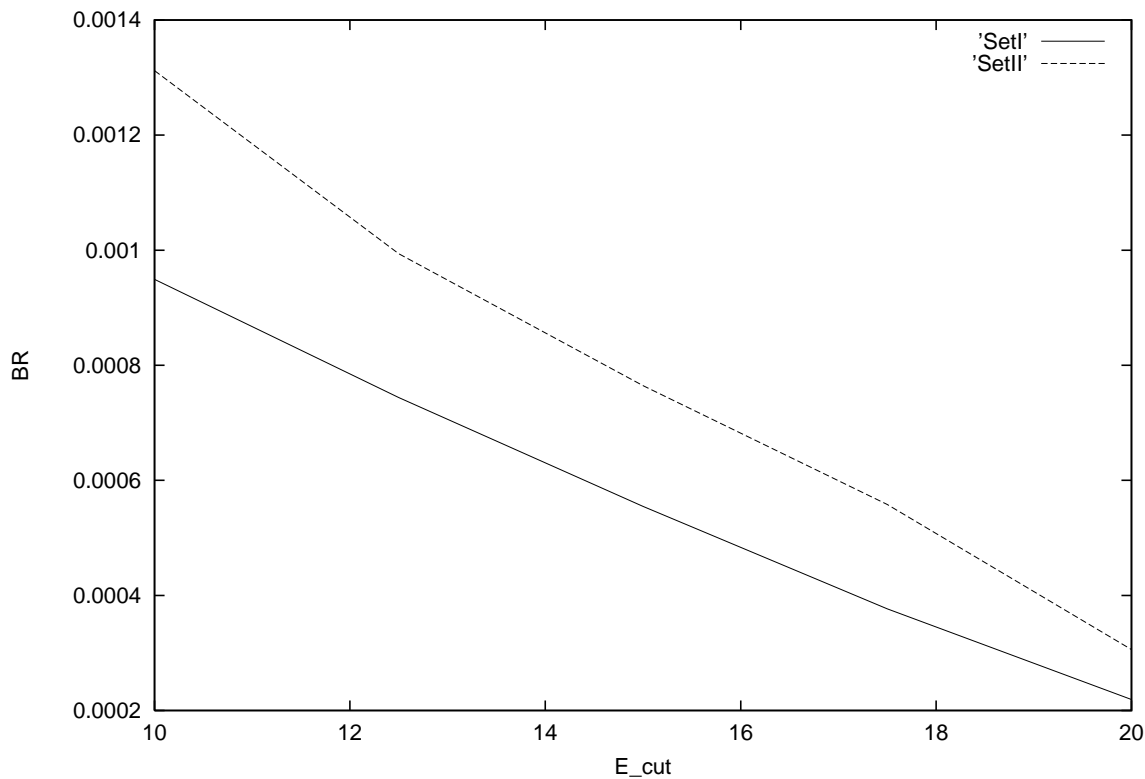}}  
\vskip 3.8in\relax\noindent
\caption{ Branching ratio into $q \bar q G$ final state times branching
fraction of the $G\rightarrow K_s K_s$ decay }
\end{figure}

Because of the good reconstruction efficiency for the $K_S$ at LEP,
we focus on the decay of the glueball into $K_S K_S$, rather
than for its decay into $\eta$'s for which the efficiency is
rather poor. Theoretical estimates \cite{decay} for the decay branching
ratio of the glueball in the $K_S K_S$ channel suggest that this
could be conservatively placed at about 2.5\%. We present our
results in terms of the branching ratio into the $q \bar q G$
final state, with the branching ratio of the glueball decay
into $K_S K_S$ also folded in. Thus we define 
\begin{equation}
{\rm BR} = {\Gamma(Z\rightarrow q \bar q G X) \over
\Gamma(Z\rightarrow q \bar q )} \cdot
{\Gamma(G\rightarrow K_s K_s) \over
\Gamma(G\rightarrow {\rm all} )} .
\end{equation}
In Fig.~2, we have shown our 
results for this branching ratio as a function of the cut on
the energy of the gluon jet denoted as $E_{\rm cut}$. These
are shown for both sets of input fragmentation functions~--
the curve marked Set I is with the pion fragmentation function
and that marked Set II is the $(1-z)^2$ fragmentation function.
Assuming  four million hadronic $Z$'s and folding in a $K_S$ 
reconstruction efficiency factor
(which is taken to be 18\%), we find that with a $E_{\rm cut}$ 
of about 15~GeV one would expect of the order of 100 events in 
the $K_S K_S$ channel, for the Set I fragmentation function.
For the Set II fragmentation function, this number varies by about
10\%. Thus it should be possible for any of the four LEP groups
to mount a glueball search on their three-jet hadronic events
from the $Z$. As mentioned earlier, we have made a rather
conservative choice for the normalisation of the fragmentation
function. If the normalisation of the exclusive distribution
amplitude for glueballs relative to that of the pion \cite{carlson}
is taken as a bench-mark, then we could expect a larger normalisation
for the fragmentation function and a correspondingly larger number
of glueball events in the hadronic decay of the $Z$.

We also find, from our computations, that the $z$ values that
are sampled in the fragmentation process lie in a not very
broad range at relatively small $z$ between 0.05 and
0.25. The lower value of $z$ accessed, is close to the
kinematic lower limit. The existence of the upper cut-off
on $z$ suggests that it may be able to improve the efficiency
of the glueball search by restricting the lab energies of the
glueball to be less than about a quarter of the energy in the
gluon jet. As mentioned earlier, we had used 
a multiplicative threshold factor in the fragmentation function.
But we find that the cuts on $d_{ij}$ and on the gluon energy
ensure that the energies involved in the fragmentation process
are large enough, so that the effect of this factor is negligible.
A similar kinematic behaviour results in the production of quarkonia
through fragmentation of high energy gluons \cite{jpsi}.

We would like to emphasise that the search we have proposed in
this paper is for a pure glueball state. It is, however, quite
likely that a glueball state in the mass range of 1.5 or 2 GeV
may mix with scalar isosinglet $q \bar q$ states in the same
mass region. Indeed, such a mixing has been invoked in the
analysis of the $f_0 (1500)$~ -- a glueball candidate. It has been
pointed out \cite{close3} that the mixing of the scalar glueball state
with a $q \bar q$ state nearly degenerate in mass can change the glueball
couplings so that the decays of the mixed state need not be such as
to give equal fraction of $\pi$'s and $K$'s, as would be expected in
the case of the decays of the pure glueball state 
\footnote{The expectation that a pure glueball state would decay into
equal fractions of $\pi$'s and $K$'s is rather naive and ignores
the decay dynamics. The lattice calculation in Ref.~\cite{svw}, in
fact, shows that these rates are unequal and, in particular, in
agreement with the experimental data on $f_J (1710)$.}. 
In the event that
the mixing is substantial, we would expect that the $K_S K_S$ branching
ratio of the mixed state to be reduced from the value used in the
present calculation by the square of the cosine of the mixing angle.

We have, in this letter, proposed a new way of exploring a glueball
in the fragmentation of hard gluon jets at LEP. Our estimated numbers
do look sufficiently encouraging for any of the four LEP experiments
to mount a glueball search in this channel. The observation of the
glueball state will provide a confirmation of one of the important
non-perturbative predictions of quantum chromodynamics.

\vskip15pt
{\it Acknowledgements: This work germinated during the Workshop on High 
Energy Physics Phenomenology V (WHEPP V) held at IUCAA, Pune, India
under the sponsorship of the S.N. Bose National Centre for Basic
Sciences. We thank Sunanda Banerjee for many helpful discussions.
We thank the referee for drawing our attention to Ref.~\cite{svw}.}

\end{document}